\documentclass{article}
\usepackage{graphicx}
\usepackage{amsmath}
\usepackage{amsfonts}
\usepackage{amssymb}

\begin{document}

\title{Non-Holonomic Control IV : Coherence Protection in a Rubidium isotope}
\author{E. Brion\\\emph{Laboratoire Aim\'{e} Cotton, }\\\emph{CNRS II, B\^{a}timent 505, }\\\emph{91405 Orsay Cedex, France.}
\and V.M. Akulin\\\emph{Laboratoire Aim\'{e} Cotton, }\\\emph{CNRS II, B\^{a}timent 505, }\\\emph{91405 Orsay Cedex, France.}
\and D. Comparat\\\emph{Laboratoire Aim\'{e} Cotton, }\\\emph{CNRS II, B\^{a}timent 505, }\\\emph{91405 Orsay Cedex, France.}
\and I. Dumer\\\emph{College of Engineering, }\\\emph{University of California, }\\\emph{Riverside, CA 92521, USA. }
\and V. Gershkovich\\\emph{Institut des Hautes Etudes Scientifiques,}\\\emph{ Bures-sur-Yvette, France. }
\and G. Harel\\\emph{Department of Computing, }\\\emph{University of Bradford, }\\\emph{Bradford, West Yorkshire BD7 1DP, United Kingdom. }
\and G. Kurizki\\\emph{Department of Chemical Physics, }\\\emph{Weizmann Institute of Science, }\\\emph{76100 Rehovot, Israel. }
\and I. Mazets\\\emph{Department of Chemical Physics, }\\\emph{Weizmann Institute of Science, }\\\emph{76100 Rehovot, Israel. }\\\emph{A.F. Ioffe Physico-Technical Institute, }\\\emph{194021 St. Petersburg, Russia. }
\and P. Pillet\\\emph{Laboratoire Aim\'{e} Cotton, }\\\emph{CNRS II, B\^{a}timent 505, }\\\emph{91405 Orsay Cedex, France.}}
\maketitle
\begin{abstract}
In this paper, we present a realistic application of the coherence protection method proposed in the previous article. A qubit of information encoded on the two spin states of a Rubidium isotope is protected from the action of electric and magnetic fields. 
\end{abstract}

\section{Introduction}

The coherence protection method presented in the previous article is applied to a Rubidium isotope, in which the information part corresponds to the spin of the exterior electron, whereas the orbital part of the wavefunction plays the role of the ancilla. 

The different steps of our protection method are illustrated on this specific example. Adding the ancilla is achieved through pumping the atom from the level 5s into the shell 60f. The coding and decoding matrices are applied through the non-holonomic control technique. Finally, the protection step is achieved through the simultaneous emissions of three photons : two emissions are stimulated, whereas the third is spontaneous and constitutes the irreversible process needed in this step. 

The paper is organized as follows. In the second section, we present the system and motivate our choice. In the third section we review each step of our method in detail and show their implementations on this specific physical configuration. Moreover, experimental problems and limitations are discussed.

\section{Presentation of the system}

In this paper, we show how to protect one qubit of information
encoded on the two spin states of the ground level 5s of the radioactive
isotope $^{78}$Rb against the action of $M=6$ error-inducing Hamiltonians
$\widehat{E}_{m}$. For numerical calculations we considered 3 magnetic
Hamiltonians
\[
\left\{  \widehat{E}_{k}^{\beta}\propto\widehat{L}_{k}+2\widehat{S}%
_{k},k=x,y,z\right\}  ,
\]
and 3 electric Hamiltonians of second order
\[
\left\{  \widehat{E}_{k,l}^{\varepsilon}\propto\widehat{r}_{k}^{2}-\widehat
{r}_{l}^{2},k,l=x,y,z,\text{ }k<l\right\}  .
\]

Let us first motivate the choice of the Rubidium atom. Alkali atoms like Rb are very interesting for
our purpose because of their hydrogen-like behavior : indeed, such an atom is the
''natural'' compound of an information subsystem, the spin part of the wavefunction,
and an ancilla, the orbital part of the quantum state. As we shall see,
one can easily increase the dimensionality of the ancilla by simply pumping the
atom towards a shell of higher orbital angular momentum $L$.
\begin{figure}
[ptb]
\begin{center}
\includegraphics[
height=2.9032in,
width=2.0254in
]
{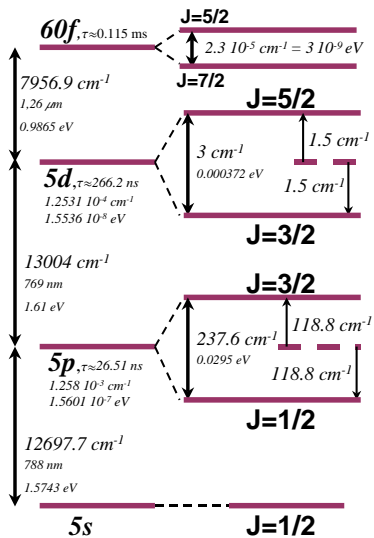}
\caption{Spectrum of $^{78}$Rb: The useful part of the spectrum of Rubidium is
represented.}
\label{Fig8}
\end{center}
\end{figure}

Among all alkali systems we chose $^{78}$Rb because of its spectroscopic
features (Fig.\ref{Fig8}): in particular, $^{78}$Rb has no hyperfine
structure (its nuclear spin is $0$) which ensures that the ground level $5s$
is degenerate (this is necessary for the projection scheme as we shall see
below). Moreover it has a long enough lifetime ($\tau\simeq17.66\min$) for the
proposed experiment.

\section{The different steps of the method}

We shall now review each step of our method in detail. The
information we want to protect is initially encoded on the two spin states
$\left|  \nu_{1}\right\rangle =\left|  5s,j=\frac{1}{2},m_{j}=-\frac{1}%
{2}\right\rangle $ and $\left|  \nu_{2}\right\rangle =\left|  5s,j=\frac{1}%
{2},m_{j}=\frac{1}{2}\right\rangle $ of the ground level $5s$ of the atom:
the two-dimensional ($I=2$) information space $\mathcal{H}_{I}=Span\left[
\left|  \nu_{1}\right\rangle ,\left|  \nu_{2}\right\rangle \right]  $ is spanned by these two states. The first step of our scheme consists in
adding an ancilla $\mathcal{A}$ to the information system. In the present setting, the role of
$\mathcal{A}$ is played by the orbital part of the wavefunction. In the ground
state ($L=0$), its dimension is $A=2L+1=1$ (roughly speaking, there is no
ancilla). If we want to protect one qubit of information against $M=6$
error-inducing Hamiltonians, we have to increase the dimensionality of the
ancilla up to $A=M+1=7$ (according to the Hamming bound presented in the previous paper), by pumping the atom towards a shell $nf$ ($L=3$). We choose the highly excited Rydberg state $60f$ so as to make the fine structure as weak as possible (the
splitting for $60f$ is approximately $10^{-5}cm^{-1}$), which shall be neglected in a first approach : thus, he $N=I\times A=2\times7=14$ basis vectors of the total Hilbert space $\mathcal{H}=\mathcal{H}_{I}\otimes\mathcal{H}_{A}$ are almost perfectly degenerate ; the
validity of this approximation will be discussed at the end of this section.
To be more specific, the pumping is done in the following way
\begin{align*}
\left|  \nu_{1}\right\rangle  &  \longrightarrow\left|  \gamma_{1}\right\rangle =\left|  60f,j=\frac{5}{2},m_{j}=-\frac{3}{2}\right\rangle \\
\left|  \nu_{2}\right\rangle  &  \longrightarrow\left|  \gamma_{2}%
\right\rangle =\left|  60f,j=\frac{5}{2},m_{j}=-\frac{1}{2}\right\rangle .
\end{align*}
In other words, using the terminology of the previous sections, the
information initially stored in $\mathcal{H}_{I}$ is transferred into
\[
\mathcal{C}=Span\left[  \left|  \gamma_{1}\right\rangle =\left|
60f,j=\frac{5}{2},m_{j}=-\frac{3}{2}\right\rangle ,\right.  \left.  \left|
\gamma_{2}\right\rangle =\left|  60f,j=\frac{5}{2},m_{j}=-\frac{1}%
{2}\right\rangle \right]  .
\]
The choice of the subspace $\mathcal{C}$ might appear arbitrary at this stage,
but it will be justified later by the practical feasibility of the projection
process onto $\mathcal{C}$. Note that $\mathcal{C}$\ is an
''entangled'' subspace, whose basis vectors $\left\{  \left|  \gamma
_{i}\right\rangle \right\}  _{i=1,2}$\ are generic entangled states of the
spin and orbital parts: this means that the projection step will not consist
in a simple measurement of the ancilla but will involve a more intricate
process we shall describe in detail later.

\begin{figure}
[ptb]
\begin{center}
\includegraphics[
height=3.5483in,
width=5.0989in
]
{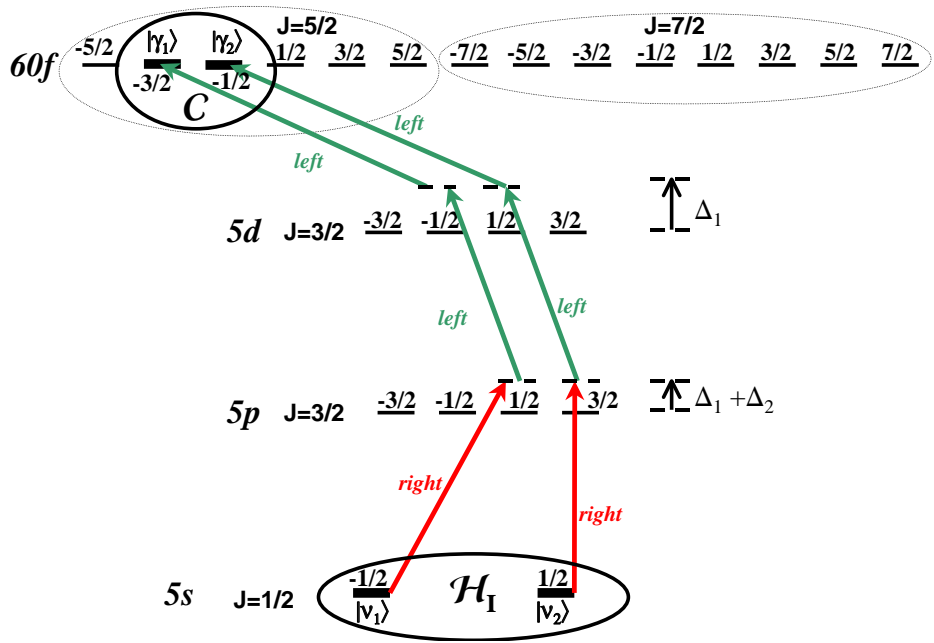}
\caption{Ancilla adding by Pumping. Photon polarization and
involved sub-Zeeman levels are represented.}
\label{Fig9}
\end{center}
\end{figure}

Practically, the pumping can be achieved as follows. Three lasers
are applied to the atom: the first laser is right polarized and slightly detuned from the
transition $\left(  5s\leftrightarrow5p\right)  $ whereas the second and third
lasers are left polarized and slightly detuned from the transitions $\left(
5p_{\frac{3}{2}}\leftrightarrow5d_{\frac{3}{2}}\right)  $\ and $\left(
5d_{\frac{3}{2}}\leftrightarrow60f\right)  $\ respectively. The role of the detunings
is to forbid real one-photon processes: thus, the atom can only absorb three photons
simultaneously and is thereby excited from the ground level $5s$ to the
Rydberg level $60f$. By using selection rules, one can construct the allowed
paths represented on Fig.\ref{Fig9}: these paths only couple $\left|  \nu
_{1}\right\rangle $ and $\left|  \nu_{2}\right\rangle $\ to $\left|
\gamma_{1}\right\rangle $ and $\left|  \gamma_{2}\right\rangle $, respectively.

\bigskip

\begin{figure}
[ptb]
\begin{center}
\includegraphics[
height=2.904in,
width=4.9917in
]
{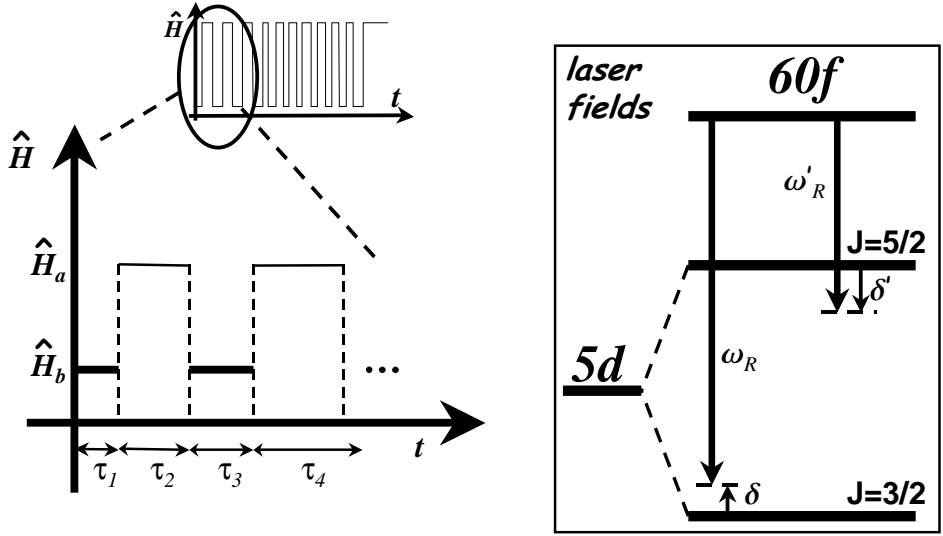}
\caption{Coding step through the non-holonomic control technique. The two
Hamiltonians $\widehat{H}_{a}$\ and $\widehat{H}_{b}$ are alternately applied
to the system during pulses of timings $\left\{  \tau_{i}(ns)\right\}
=$\{3.9763, 6.4748, 4.2274, 3.6259, 2.8717, 3.6281, 7.2263, 6.4260, 4.8070,
5.0394, 6.5242, 4.8890, 4.2400, 7.3834, 4.8653, 5.4799, 4.5341, 4.3099,
6.2959, 3.7346, 6.5293, 6.8586, 6.0749, 5.1213, 4.6806, 3.4985, 3.9909,
4.6701, 4.5168, 6.4702, 4.7787, 5.3476, 3.4567, 3.8009\}. The frequencies of
the laser fields involved in the encoding step are represented on the spectrum
of the Rubidium atom. The fine structure of the Rydberg level $60f$ is not
represented.}%
\label{Fig10}
\end{center}
\end{figure}

The second step consists in imposing the coding matrix to the system by the non-holonomic
control technique: to this end, we submit the atom to $n_{C}=34$ control pulses of timings $\left\{  t_{i}\right\}
_{i=1,...,34}$, during which two different combinations of magnetic and Raman
electric Hamiltonians are alternately applied (see Fig.\ref{Fig10}). To be more
explicit, during odd-numbered pulses (''A'' type pulses) we apply the constant
magnetic field
\[
\overrightarrow{B}=\left(
\begin{array}
[c]{c}%
B_{x}=7\;10^{-3}T\\
B_{y}=8.2\;10^{-3}T\\
B_{z}=-6.8\;10^{-3}T
\end{array}
\right)
\]
associated with the Zeeman Hamiltonian $\widehat{W}_{Z}$, and two sinusoidal electric laser fields
\begin{align*}
\overrightarrow{E}_{a}(t) &  =\operatorname{Re}\left[  \underline
{\overrightarrow{E}}_{a}e^{-i\omega_{R}t}\right]  ,\overrightarrow{E}%
_{a}^{\prime}(t)=\operatorname{Re}\left[  \underline{\overrightarrow{E}}%
_{a}^{\prime}e^{-i\omega_{R}^{\prime}t}\right]  ,\\
\underline{\overrightarrow{E}}_{a} &  =\left|
\begin{array}
[c]{c}%
E_{x,a}\\
E_{y,a}e^{-i\varphi_{y,a}}\\
0
\end{array}
\right.  ,\underline{\overrightarrow{E}}_{a}^{\prime}=\left|
\begin{array}
[c]{c}%
E_{x,a}^{\prime}\\
E_{y,a}^{\prime}e^{-i\varphi_{y,a}^{\prime}}\\
0
\end{array}
\right.  ,
\end{align*}
whose frequencies $\omega_{R}$ and $\omega_{R}^{\prime}$\ are respectively
slightly detuned from the two transitions $\left(  60f\leftrightarrow
5d,j=\frac{3}{2}\right)  $ and $\left(  60f\leftrightarrow5d,j=\frac{5}
{2}\right)  $ (detunings $\delta$ and $\delta^{\prime}$). The characteristic
values of these fields are
\begin{align*}
E_{x,a} &  =E_{x,a}^{\prime}=8.5\;10^{5}V.m^{-1}\\
E_{y,a} &  =E_{y,a}^{\prime}=5.2\;10^{6}V.m^{-1}\\
\varphi_{y,a} &  =\varphi_{y,a}^{\prime}=2.3\\
\hbar\omega_{R} &  =0.986324\;eV=7955.14\;cm^{-1}\\
\delta &  =-0.000010\;eV=-0.080654\;cm^{-1}\\
\hbar\omega_{R}^{\prime} &  =0.986676\;eV=7958.14\;cm^{-1}\\
\delta^{\prime} &  =0.000010\;eV=0.080654\;cm^{-1}.
\end{align*}
The intensity of the laser beams are typically of the order of $2$
$10^{8}W.cm^{-1}$. The Raman Hamiltonian associated with these fields is
denoted by $\widehat{W}_{R,A}$ and the total perturbation is $\widehat{P}_{a}=\widehat{W}_{Z}+\widehat{W}_{R,A}$. During even-numbered pulses (''B''
type pulses), we apply the same magnetic field as for A type pulses, which is
experimentally convenient, and two sinusoidal electric laser fields
\begin{align*}
\overrightarrow{E}_{b}(t) &  =\mbox{Re}\left[  \underline{\overrightarrow{E}%
}_{b}e^{-i\omega_{R}t}\right]  ,\overrightarrow{E}_{b}(t)=\mbox{Re}\left[
\underline{\overrightarrow{E}}_{b}^{\prime}e^{-i\omega_{R}^{\prime}t}\right]
,\\
\mbox{where }\underline{\overrightarrow{E}}_{b} &  =\left|
\begin{array}
[c]{c}%
E_{x,b}\\
E_{y,b}e^{-i\varphi_{y,b}}\\
0
\end{array}
\right.  ,\underline{\overrightarrow{E}}_{b}^{\prime}=\left|
\begin{array}
[c]{c}%
E_{x,b}^{\prime}\\
E_{y,b}^{\prime}e^{-i\varphi_{y,b}^{\prime}}\\
0
\end{array}
\right.  ,
\end{align*}
whose frequencies are the same as above and whose characteristic values are
\begin{align*}
E_{x,b} &  =E_{x,b}^{\prime}=-5.2\text{ }10^{6}V.m^{-1}\\
E_{y,b} &  =E_{y,b}^{\prime}=8.5\;10^{5}V.m^{-1}\\
\varphi_{y,a} &  =\varphi_{y,a}^{\prime}=2.3.
\end{align*}
The Raman Hamiltonian associated with these fields is denoted by $\widehat
{W}_{R,B}$. The corresponding perturbation is $\widehat{P}_{b}=\widehat{W}%
_{Z}+\widehat{W}_{R,B}$. As the fine structure of the level
$60f$ is neglected, the unperturbed Hamiltonian $\widehat{H}_{0}$ is $0$ and the total Hamiltonian has the following form: $\widehat{H}_{a}=\widehat{P}_{a}$ during ''A'' pulses, $\widehat{H}_{b}=\widehat{P}_{b}$ during ''B'' pulses. The $34$ different timings have been calculated so that
\[
\widehat{U}(\tau_{1},...,\tau_{34})=e^{-i\widehat{H}_{B}\tau_{34}}e^{-i\widehat
{H}_{A}\tau_{33}}\ldots e^{-i\widehat{H}_{A}\tau_{1}}=\widehat{C}\]
checks the correction conditions presented in the previous paper. At the end of the coding step, the information
is transferred into the code space $\widetilde{\mathcal{C}}=\widehat
{C}\mathcal{C}$, encoded on the codewords $\left\{  \left|  \widetilde{\gamma
}_{i}\right\rangle =\widehat{C}\left|  \gamma_{i}\right\rangle \right\}
_{i=1,2}$.

As one can see on Fig.\ref{Fig10}, the total duration of a control
period ($\simeq125ns$) is approximately $10^{3}$ times shorter than the
lifetime of $60f$ Rydberg state which is approximately $0.115ms$, and the different pulse timings range between $2.9ns$ and $7.4ns$, which are feasible.

\bigskip

\begin{figure}
[ptb]
\begin{center}
\includegraphics[
height=2.6377in,
width=4.3664in
]
{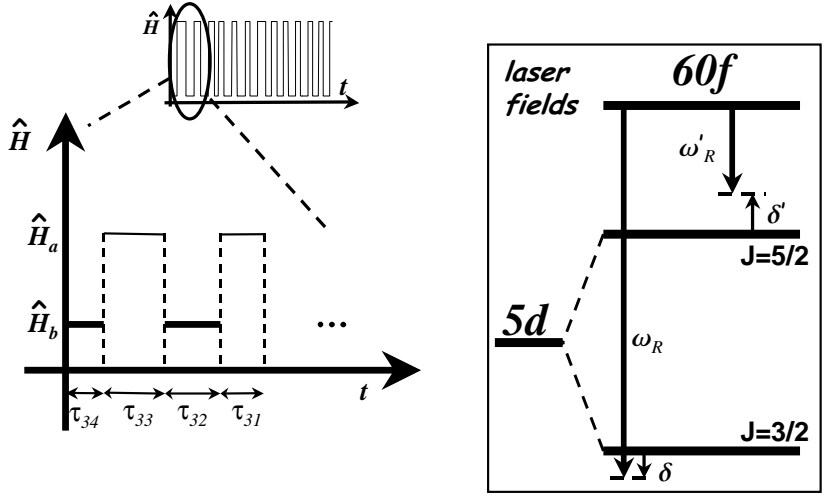}
\caption{Decoding step through the non-holonomic control technique. We reverse
the magnetic field and the detunings of electric fields, as represented on the
spectrum of the Rubidium atom, and apply the same control sequence as for
coding (same timings) in the reverse way.}
\label{Fig11}
\end{center}
\end{figure}

After a short time, due to the action of the error Hamiltonians, the information stored in the system acquires a small erroneous component, which is orthogonal to the code space $\widetilde{\mathcal{C}}$. Then, we decode the information through the application of the matrix $\widehat{C}^{-1}$. To this end, we reverse $\overrightarrow{B}$ and the detunings $\delta$ and
$\delta^{\prime}$, while leaving all the other values unchanged (this amounts to
taking the opposite of Hamiltonians $\widehat{H}_{a}$ and $\widehat{H}_{b}$),
and apply the same sequence of control pulses backwards: we start with an
''A'' pulse whose timing is $\tau_{34}$, then apply a ''B'' pulse during
$\tau_{33},$ etc. (see Fig.\ref{Fig11}). The decoding step yields an
erroneous state whose projection onto $\mathcal{C}$ is the initial information state.

\begin{figure*}
[ptb]
\begin{center}
\includegraphics[
height=3.544in,
width=5.1162in
]
{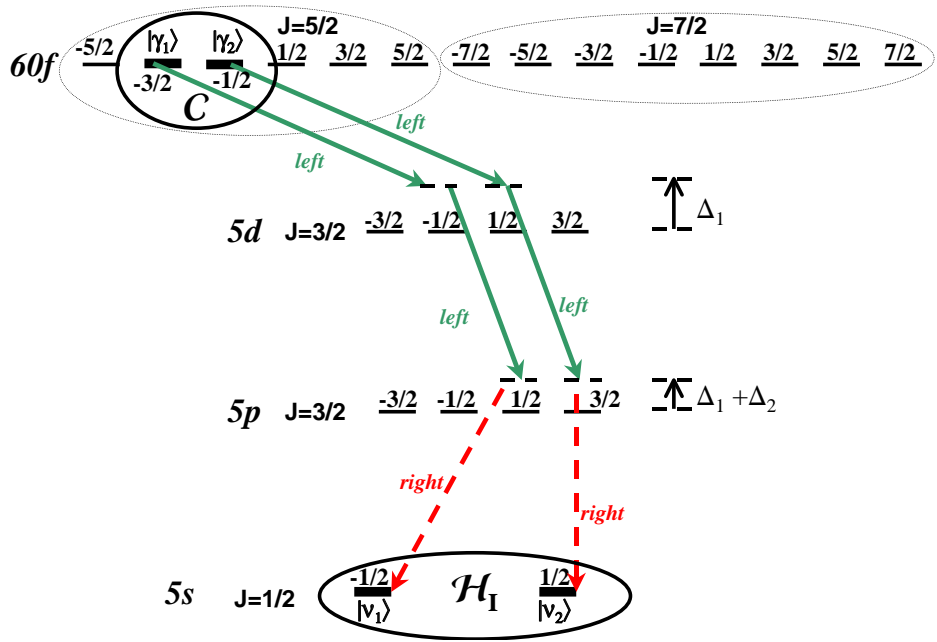}
\caption{Projection path. The lasers involved are marked by solid arrows, 
the spontaneous photon is represented by a dashed arrow. The
different polarizations are specified. The fine structure of the level $60f$ is not represented.}%
\label{Fig12}
\end{center}
\end{figure*}

In the last step, we project the erroneous state vector onto the subspace
$\mathcal{C}$ to recover the initial information. Projection is a non-unitary
process which cannot be achieved through a Hamiltonian process, but requires
the introduction of irreversibility. To this end, we make use of a path which
is symmetric with the pumping step, consisting in two stimulated and one
spontaneous emissions. To be more explicit, we apply two left circularly
polarized lasers (see Fig.\ref{Fig12}) slightly detuned from the transitions
$\left(  60f\longleftrightarrow5d,j=\frac{3}{2}\right)  $ and $\left(
5d,j=\frac{3}{2}\longleftrightarrow5p,j=\frac{3}{2}\right)  $. Due to these
laser fields, the atom is likely to fall towards the ground state and emit two
stimulated and one spontaneous photons.

If a circularly right-polarized spontaneous photon is emitted, the selection rules show that the only states to be coupled to the ground level are $\left|  \gamma_{1}\right\rangle $ and $\left|  \gamma
_{2}\right\rangle $ to $\left|  \nu_{1}\right\rangle $ and $\left|  \nu
_{2}\right\rangle $, respectively (see Fig.\ref{Fig12}). In other terms, the emission of such a spontaneous photon brings the ''correct'' part of the state vector back into $\mathcal{H}_{I}=Span\left[  \left|  \nu
_{1}\right\rangle ,\left|  \nu_{2}\right\rangle \right]  $. On the contrary,
the other cases - ''left-polarized'', ''linearly-polarized spontaneous
photon'', or ''no photon at all''- do not lead to the right projection
process.

The ''left-polarized photon'' and ''no photon emitted'' cases are quite
unlikely, since their probability is proportional to the square of the error amplitude, that is to the square of the very short Zeno interval. By contrast, the ''linearly polarized photon'' case
is quite annoying because it mixes the two paths $\left|  \gamma
_{1}\right\rangle \longrightarrow\left|  \nu_{1}\right\rangle $ and $\left|
\gamma_{2}\right\rangle \longrightarrow\left|  \nu_{2}\right\rangle $ : one has to get rid of this
parasitic process by minimizing its relative probability with respect to the process followed by the ''right-polarized'' photon emission. 
This can be achieved by launching the $^{78}$Rb atom, previously cooled, into a Fabry-Perot cavity, in an atomic fountain manner (fine tuning of the lasers driving the $60f-5d$ and $5d-5p$ transition will be
necessary to avoid reflection of the external laser radiation from the
cavity). The decay rate for the 3-photon transition $\left|  \gamma
_{i}\right\rangle \longrightarrow\left|  \nu_{i}\right\rangle $ is
\begin{displaymath}
\Gamma_{\gamma_{i}\nu_{i}}=2\pi\left|  \frac{d_{\gamma_{i}\lambda_{j}}E_{1}%
}{\hbar\Delta_{1}}\right|  ^{2}\left|  \frac{d_{\lambda_{j}\mu_{k}}E_{2}%
}{\hbar(\Delta_{1}+\Delta_{2})}\right|  ^{2}\overline{2\pi\hbar ck_{s}\left|
\overrightarrow{d}_{\mu_{k}\nu_{i}}\overrightarrow{e}_{R}^{\ast}\right|
^{2}\varrho\left(  \overrightarrow{k}_{s}\right)  },
\end{displaymath}
where $\overrightarrow{k}_{s}$ denotes the wave vector of the spontaneously emitted
photon, $\overrightarrow{e}_{R}$ is the left-polarized photon polarization
unit vector, $\varrho\left(  \overrightarrow{k}_{s}\right)  $ is the density
of states (normalized to the cavity volume) for the cavity field at
$\overrightarrow{k}_{s}$, and the bar denotes averaging over the directions of
$\overrightarrow{k}_{s}$. The transition dipole moments are denoted by
$d_{ab}$: during the projective process the states coupled to $\left|
\gamma_{1}\right\rangle $ and $\left|  \gamma_{2}\right\rangle $ are 
\begin{eqnarray*}
\left\{  \left|  \lambda_{1}\right\rangle =\left|  5d,j=\frac
{3}{2},m_{j}=-\frac{1}{2}\right\rangle , \left|  \lambda_{2}\right\rangle
=\left|  5p,j=\frac{3}{2},m_{j}=+\frac{1}{2}\right\rangle \right\}, \\ \left\{  \left|  \mu_{1}\right\rangle =\left|  5d,j=\frac{3}{2},m_{j}%
=\frac{1}{2}\right\rangle , \left|  \mu_{2}\right\rangle =\left|  5p,j=\frac
{3}{2},m_{j}=\frac{3}{2}\right\rangle \right\}.
\end{eqnarray*} 
respectively.

The presence of the cavity enhances the density of states for the modes propagating
paraxially to the $z$-axis and ensures that 
\[
\Gamma_{\gamma_{1}\nu_{1}},\Gamma_{\gamma_{2}\nu_{2}}\gg\left|  \frac
{d_{\gamma_{i}\lambda_{j}}E_{1}}{\hbar\Delta_{1}}\right|  ^{2}\left|
\frac{d_{\lambda_{j}\mu_{k}}E_{2}}{\hbar(\Delta_{1}+\Delta_{2})}\right|
^{2}\gamma,
\]
where $\gamma$ is the decay rate of $\left|  5p,j=\frac{3}{2},m_{j}=+\frac{1}{2}\right\rangle $ into $\left|  5s,j=\frac{1}{2},m_{j}=+\frac{1}{2}\right\rangle $, so that the undesired process followed by the
$\pi$-photon emission is relatively less important than it were in free space. The dynamics of the density matrix elements $\rho_{ab}$ is governed by the following system ($i=0,1$):
\begin{eqnarray}
\dot{\rho}_{\gamma_{i}\gamma_{i}}=-\Gamma_{\gamma_{i}\nu_{i}}\dot{\rho
}_{\gamma_{i}\gamma_{i}}, \nonumber \\
\dot{\rho}_{\nu_{i}\nu_{i}}=\Gamma_{\gamma_{i}\nu_{i}}\dot{\rho}%
_{\gamma_{i}\gamma_{i}}, \nonumber \\
\dot{\rho}_{\gamma_{1}\gamma_{2}}=-\frac{1}{2}(\Gamma_{\gamma_{1}\nu_{1}%
}+\Gamma_{\gamma_{2}\nu_{2}})\rho_{\gamma_{1}\gamma_{2}}, \nonumber \\
\dot{\rho}_{\nu_{1}\nu_{2}}=\sqrt{\Gamma_{\gamma_{1}\nu_{1}}%
\Gamma_{\gamma_{2}\nu_{2}}}\rho_{\gamma_{1}\gamma_{2}}. \nonumber 
\end{eqnarray}
To avoid dephasing which would corrupt the information, the
coherence matrix element $\rho_{\gamma_{1}\gamma_{2}}$ must be transferred
with the maximum efficiency $\eta$ 
\[
\eta=\frac{2\sqrt{\Gamma_{\gamma_{1}\nu_{1}}\Gamma_{\gamma_{2}\nu_{2}}}%
}{\Gamma_{\gamma_{1}\nu_{1}}+\Gamma_{\gamma_{2}\nu_{2}}}%
\]
into $\rho_{\nu_{1}\nu_{2}}$. According to the Wigner-Eckart theorem, we have
\[
\frac{\Gamma_{\gamma_{1}\nu_{1}}}{\Gamma_{\gamma_{2}\nu_{2}}}=\left(
\frac{C_{3/2~-1/2~1~-1}^{5/2~-3/2}C_{3/2~1/2~1~-1}^{3/2~-1/2}C_{1/2~-1/2~1~1}%
^{3/2~1/2}}{C_{3/2~1/2~1~-1}^{5/2~-1/2}C_{3/2~3/2~1~-1}^{3/2~1/2}%
C_{1/2~1/2~1~1}^{3/2~3/2}}\right)  ^{2},
\]
whence $\eta=12\sqrt
{2}/17\approx0.99827$. In other words, the probability of error during the Zeno
projection stage due to the small difference of the Clebsch-Gordan coefficient
products for the two paths is equal to or less than $1-\eta\approx0.00173$
(the equality is reached if the initial state is $(\left|  0\right\rangle
\pm\left|  1\right\rangle )/\sqrt{2}$). Note that the states $60f$,
$5d$, and $5p$ have finite lifetimes $\tau_{k}$ (see Fig.\ref{Fig8}). Thus the
transition rates $\Gamma_{\gamma_{i}\nu_{i}}$ must be much larger than
$1/\tau_{60f}$, $\left|  \frac{d_{\gamma_{i}\lambda_{j}}E_{1}}{\hbar\Delta
_{1}}\right|  ^{2}/\tau_{5d}$, and $\left|  \frac{d_{\gamma_{i}\lambda_{j}}E_{1}}{\hbar\Delta_{1}}\right|  ^{2}\left|  \frac{d_{\lambda_{j}\mu_{k}}E_{2}}{\hbar(\Delta_{1}+\Delta_{2})}\right|  ^{2}/\tau_{5p}$, in order to
minimize errors caused by the decay of these unstable states.

To complete the projection step, one has now to transfer the atom back into its coherent
superposition on the shell $60f$: this is achieved by the same pumping
sequence as in the first step. The mismatch of the Clebsch-Gordan coefficient
products will cause again the error probability $1-\eta$. The information is
then restored with very high probability and the system is ready to undergo a
new protection cycle.

\bigskip

Until now, we have neglected the fine structure splitting of the level $60f$, which is approximately $2.10^{-5}cm^{-1}$ and corresponds to a period $\tau_{f}\sim1.5\mu s$. To conclude this section, we
shall now take it into account and see how it affects each step of our scheme.

Obviously the pumping and projection steps will not be affected by the fine
structure, since the information-carrying vectors $\left\{  \left|
\gamma_{1}\right\rangle ,\left|  \gamma_{2}\right\rangle \right\}  $ belong to
the same multiplet $\left(  J=5/2\right)$.

The coding and decoding steps are neither modified by the existence of the
fine structure. Since the typical period of the fine structure
Hamiltonian $\tau_{f}\sim1.5\mu s$\ is more than $10$ times longer than the
total duration of the coding or decoding steps, it is legitimate to neglect
its effect.

The influence of the fine structure on the free evolution period during which
errors are likely to occur is more complicated to study in the general case.
However, two simple limiting regimes can be considered. If the spectrum of the
coupling functions $f_{m}(t)$'s is very narrow (\emph{i.e.} if the variation
timescale of the $f_{m}(t)$'s is much longer than $\tau_{f}$), one can show
that our scheme applies directly as though there were no fine structure,
provided the error Hamiltonians $\left\{  \widehat{E}_{m}\right\}  $ are
replaced by $\left\{  \widehat{E}_{m}^{(0)}\right\}  $, where $\widehat{E}_{m}^{(0)}$ is obtained from $\widehat{E}_{m}$\ by simply setting to zero the
rectangular submatrices which couple the two multiplets $\left(
J=5/2,7/2\right)  $. The second limiting regime corresponds to a very broad
spectrum for the $f_{m}(t)$'s (variation timescale much shorter than $\tau
_{f}$): in that case, one can show that our scheme applies provided one
chooses a Zeno interval multiple of $\tau_{f}$.

Finally, it must be emphasized that Rydberg atoms, though long-lived are also very sensitive to collisional processes as well as Doppler effects, which result in a fast coherence loss ; nevertheless, we hope that in a single (or few) atom experiment using cold atoms the coherence lifetime can be extended to hundreds of $ns$. These effects have been omitted in all this section in which we just intended to provide a pedagogical demonstration of our method on a simple physical system ; but an experimentally feasible setup should obviously deal with these unavoidable drawbacks of Rydberg states.

\section{Conclusion}

In this paper, a realistic application of our coherence protection method has been proposed : it has been shown that, in principle, one qubit of information encoded on the spin states of a Rubidium isotope can be protected from the action of parasitic electric and magnetic fields. The different steps of our technique can be implemented on this specific example : adding the ancilla is achieved through pumping ; information is coded and decoded through non-holonomic control ; projection is achieved by a three-photon process, involving spontaneous emission.

Practical feasibility of our scheme has been discussed, and experimental problems have been raised. However relevant, these limitations do not restrict the applicability of our method, and the pedagogical example considered here demonstrates its implementability. 

\end{document}